\documentclass[12pt]{article}

\expandafter\ifx\csname mathrm\endcsname\relax\def\mathrm#1{{\rm #1}}\fi

\makeatletter
\if@twoside \oddsidemargin -17pt \evensidemargin 00pt
\else \oddsidemargin 00pt \evensidemargin 00pt
\fi
\expandafter\ifx\csname mathrm\endcsname\relax\def\mathrm#1{{\rm #1}}\fi

\makeatother

\def\beq{\begin{equation}}
\def\eeq{\end{equation}}
\def\beqar{\begin{eqnarray}}
\def\eeqar{\end{eqnarray}}
\def\barr#1{\begin{array}{#1}}
\def\earr{\end{array}}
\def\bfi{\begin{figure}}
\def\efi{\end{figure}}
\def\btab{\begin{table}}
\def\etab{\end{table}}
\def\bce{\begin{center}}
\def\ece{\end{center}}
\def\nn{\nonumber}

\def\text{\textstyle}

\def\de{\delta}
\def\eps{\varepsilon}

\def\si{\sigma}

\def\Ga{\Gamma}
\def\De{\Delta}

\def\refeq#1{\mbox{(\ref{#1})}}
\def\reffi#1{\mbox{Fig.~\ref{#1}}}
\def\reffis#1{\mbox{Figs.~\ref{#1}}}
\def\refta#1{\mbox{Tab.~\ref{#1}}}
\def\citere#1{\mbox{Ref.~\cite{#1}}}
\def\citeres#1{\mbox{Refs.~\cite{#1}}}

\def\solid{\raise.9mm\hbox{\protect\rule{1.1cm}{.2mm}}}
\def\dash{\raise.9mm\hbox{\protect\rule{2mm}{.2mm}}\hspace*{1mm}}

\newcommand{\GeV}{\unskip\,\mathrm{GeV}}
\newcommand{\MeV}{\unskip\,\mathrm{MeV}}
\newcommand{\TeV}{\unskip\,\mathrm{TeV}}

\def\mathswitchr#1{\relax\ifmmode{\mathrm{#1}}\else$\mathrm{#1}$\fi}

\newcommand{\PW}{\mathswitchr W}
\newcommand{\PZ}{\mathswitchr Z}

\newcommand{\PH}{\mathswitchr H}

\newcommand{\Pb}{\mathswitchr b}
\newcommand{\Pc}{\mathswitchr c}
\newcommand{\Pt}{\mathswitchr t}

\newcommand{\PWpm}{\mathswitchr {W^\pm}}

\newcommand{\Zbb}{$\PZ\to\Pb\bar\Pb$}

\newcommand{\Rb}{R_\Pb}
\newcommand{\Rc}{R_\Pc}
\newcommand{\Gb}{\Ga_\Pb}
\newcommand{\Gc}{\Ga_\Pc}

\newcommand{\Gh}{\Ga_{\mathrm h}}
\newcommand{\GT}{\Ga_{\mathrm T}}
\newcommand{\Gl}{\Ga_{\mathrm l}}

\def\mathswitch#1{\relax\ifmmode#1\else$#1$\fi}

\newcommand{\MW}{\mathswitch {M_\PW}}

\newcommand{\MZ}{\mathswitch {M_\PZ}}
\newcommand{\MH}{\mathswitch {M_\PH}}

\newcommand{\Mb}{\mathswitch {m_\Pb}}
\newcommand{\Mt}{\mathswitch {m_\Pt}}

\newcommand{\scrs}{\scriptscriptstyle}

\newcommand{\swbar}{\mathswitch {\bar s_{\scrs\PW}}}

\newcommand{\GF}{\mathswitch {G_\mu}}

\newcommand{\chidof}{\chi^2_{\mathrm{min}}/_{\mathrm{d.o.f.}}}

\newcommand{\yb}{y_\Pb}

\newcommand{\alpz}{\alpha(\MZ^2)}

\newcommand{\alpsz}{\alpha_{\mathrm s}(\MZ^2)}

\newcommand{\bos}{{\mathrm{bos}}}
\newcommand{\fer}{{\mathrm{ferm}}}

\newcommand{\SC}{{\mathrm{SC}}}

\newcommand{\yh}{y_{\mathrm h}}

\newcommand{\LEP}{{\mathrm{LEP}}}
\newcommand{\SLD}{{\mathrm{SLD}}}

\def\draftdate{\relax}
\def\mda{\relax}
\def\mua{\relax}
\def\mla{\relax}
\def\draft{
\def\thtystars{******************************}
\def\sixtystars{\thtystars\thtystars}
\typeout{}
\typeout{\sixtystars**}
\typeout{* Draft mode!
         For final version remove \protect\draft\space in source file *}
\typeout{\sixtystars**}
\typeout{}
\def\draftdate{\today}
\def\mua{\marginpar[\boldmath\hfil$\uparrow$]%
                   {\boldmath$\uparrow$\hfil}%
                    \typeout{marginpar: $\uparrow$}\ignorespaces}
\def\mda{\marginpar[\boldmath\hfil$\downarrow$]%
                   {\boldmath$\downarrow$\hfil}%
                    \typeout{marginpar: $\downarrow$}\ignorespaces}
\def\mla{\marginpar[\boldmath\hfil$\rightarrow$]%
                   {\boldmath$\leftarrow $\hfil}%
                    \typeout{marginpar: $\leftrightarrow$}\ignorespaces}
\def\Mua{\marginpar[\boldmath\hfil$\Uparrow$]%
                   {\boldmath$\Uparrow$\hfil}%
                    \typeout{marginpar: $\Uparrow$}\ignorespaces}
\def\Mda{\marginpar[\boldmath\hfil$\Downarrow$]%
                   {\boldmath$\Downarrow$\hfil}%
                    \typeout{marginpar: $\Downarrow$}\ignorespaces}
\def\Mla{\marginpar[\boldmath\hfil$\Rightarrow$]%
                   {\boldmath$\Leftarrow $\hfil}%
                    \typeout{marginpar: $\Leftrightarrow$}\ignorespaces}
\overfullrule 5pt
\oddsidemargin -15mm
\marginparwidth 29mm
}

\makeatletter

\def\eqnarray{\stepcounter{equation}\let\@currentlabel=\theequation
\global\@eqnswtrue
\global\@eqcnt\z@\tabskip\@centering\let\\=\@eqncr
$$\halign to \displaywidth\bgroup\hskip\@centering
  $\displaystyle\tabskip\z@{##}$\@eqnsel&\global\@eqcnt\@ne
  \hskip 2\arraycolsep \hfil${##}$\hfil
  &\global\@eqcnt\tw@ \hskip 2\arraycolsep $\displaystyle\tabskip\z@{##}$\hfil
   \tabskip\@centering&\llap{##}\tabskip\z@\cr}
\def\appendix{\par
 \setcounter{section}{0} \setcounter{subsection}{0}
 \def\thesection{\Alph{section}}}

\makeatother

\newcommand{\lsim}
{\;\raisebox{-.3em}{$\stackrel{\displaystyle <}{\sim}$}\;}

\begin{document}
\thispagestyle{empty}
\def\thefootnote{\fnsymbol{footnote}}
\setcounter{footnote}{1}
\null
\hfill BI-TP 96/44 \\
\null
\hfill hep-ph/9609488
\vskip 1.5cm
\begin{center}
{\Large \bf 
Implications of the Electroweak Precision Data: \\[1em]
a 1996 Update
}%
\footnote{Partially supported by the EC-network contract CHRX-CT94-0579.}
\vskip 3.0em
{\large S.\ Dittmaier and D.\ Schildknecht 
}
\vskip .5em
{\it Fakult\"at f\"ur Physik, Universit\"at Bielefeld, Germany}
\vskip 2em
\end{center} \par
\vskip 2.5cm
\vfil
{\bf Abstract} \par
The most reliable prediction for the Higgs-boson mass,
$\MH$, is obtained in a fit to the leptonic Z-resonance observables
$\Gl$ and $\swbar^2$ combined with the W-boson mass, $\MW$, and
top-quark mass, $\Mt$, measurements.
The corresponding bounds on $\MH$ are independent of potential 
uncertainties related to $\Rb$, $\Rc$, and $\alpsz$, and they are not
significantly further improved by including also the experimental
information on the inclusive (hadronic and total) Z-boson decays.
At the $1\si$ level, we obtain 
$\MH\lsim 360\GeV$ using $\swbar^2(\LEP+\SLD)$
and $\MH\lsim 540\GeV$ using $\swbar^2(\LEP)$.
Our analysis in terms of effective parameters confirms previous
conclusions with increased accuracy. 
In the mass parameter, $\De x$, and the mixing parameter, $\eps$, pure
fermion loops are sufficient, while for the coupling parameter, $\De y$,
($\MH$-insensitive) bosonic contributions are essential
for consistency with experiment, thus 
providing indirect empirical evidence for
the non-Abelian structure of the theory.
\par
\vskip 2cm
\noindent
September 1996
\null
\setcounter{page}{0}
\clearpage
\def\thefootnote{\arabic{footnote}}
\setcounter{footnote}{0}

Employing the most recent 1996 electroweak precision 
data~\cite{LEPEWWG9602}, 
in the present note
we update the previous 
analyses~\cite{zph4,zph1,zph2} 
with respect to
\renewcommand{\labelenumi}{(\roman{enumi})}
\begin{enumerate}
\item
their implications on constraining the mass $\MH$ of the Higgs scalar
within the Standard Model (SM),
in comparison with previous results \cite{mo94,el95},
\item
their implications in terms of the effective parameters 
$\De x$, $\De y$, $\eps$, and $\De\yb$, which quantify
deviations from custodial SU(2) and SU(2)$\times$U(1) symmetry
within an effective Lagrangian~\cite{zph1,zph2,kn91,bi93} for electroweak 
interactions at the Z-boson resonance
and are closely related to the
$\eps_i$ parameters of \citere{al93}.
\end{enumerate}

\section*{Bounds on the Higgs-boson mass}

As discussed in a recent analysis~\cite{zph4}, the 1995 electroweak 
precision data~\cite{LEPEWWG9502} lead to the following $1\si$ bounds 
for the mass of the Higgs boson:
\beq
\begin{array}[b]{rcl}
\MH &=& 148^{+263}_{-103}\GeV \qquad \mbox{using} \qquad
\parbox{7cm}{\hfill$\swbar^2(\LEP+\SLD)|_{'95}=0.23143\pm 0.00028,$} 
\\
\MH &=& 343^{+523}_{-219}\GeV \qquad \mbox{using} \qquad
\parbox{7cm}{\hfill$\swbar^2(\LEP)|_{'95}=0.23186\pm 0.00034.$}
\end{array}
\label{eq:MH95}
\eeq
These values and the corresponding upper bounds of 
$\MH\lsim 400\GeV$ and
$\MH\lsim 900\GeV$ at the $1\si$ level are considerably higher
than some of the results which appeared in the literature 
(compare e.g.\ \citere{el95}).
The difference between \refeq{eq:MH95} and 
results obtained by some
other authors is essentially due to the exclusion of the 1995 value
of $\Rb=\Gamma(\PZ\to\Pb\bar\Pb)/\Gamma(\PZ\to\mathrm{hadrons})$
in the four-parameter $(\Mt,\MH,\alpz,\alpsz)$
fits which leads to the results \refeq{eq:MH95}.
In fact, it was argued in \citere{zph4} that the exclusion of the 
experimental value for $\Rb$ from the set of data used to deduce the 
Higgs-boson mass was necessary in view of the upward shift of
$\Rb^{\exp}|_{'95}=0.2219\pm 0.0017$ by more than three standard deviations 
with respect to the SM prediction.
Indeed,
inclusion of the 1995 value of $\Rb^{\exp}$ in the set of data being fitted,
\begin{enumerate}
\item 
within the SM effectively amounts to imposing a
value of the top-quark mass, $\Mt$, which lies substantially below%
\footnote{Not including the direct Tevatron measurement of $\Mt$ in the 
set of data used in the fit, using $\swbar^2(\LEP+\SLD)$, we obtained 
\cite{zph4}
$\Mt=153\pm 11\GeV$ compared with 
$\Mt^{\exp}|_{'95}=180\pm 12\GeV$ \cite{mt95} from the Tevatron
measurement of 1995. In the same fit, we obtained $\MH=35^{+50}_{-18}\GeV$.}
the result of the direct measurements at the Tevatron.
This is a consequence of the enhancement of the 1995 value
of $\Rb^{\exp}$ in conjunction with the increase of the 
($\MH$-insensitive)
theoretical prediction for $\Rb$ with decreasing $\Mt$.
It is precisely this preference of a low
value of $\Mt$ which in turn implies a low value for $\MH$ due to
the well-known $(\Mt,\MH)$ correlation in the other
observables entering the fit. Since the low value of $\Mt$ is at variance
with the direct Tevatron measurement,
also the low value of $\MH$ resulting from this fit 
was rejected.
\item
allowing for a non-standard \Zbb\ vertex parametrized
by an adjustable constant, yields a value of $\Mt$ consistent with
the Tevatron measurement. At the same time, however, the stringent
upper bounds on $\MH$ resulting from including $\Rb^{\exp}$ in the
SM fits are lost. The results for $M_H$ in this
non-standard fit approximately coincide with the results \refeq{eq:MH95},
obtained by excluding $\Rb^{\exp}$ from the data set in the fit based on
the unmodified SM.
\end{enumerate}
Accordingly, in \citere{zph4} it was concluded that the
values \refeq{eq:MH95} were indeed the only 
reliable ones obtainable from the 1995 data.

The most pronounced change in the 1996 data \cite{LEPEWWG9602}
relative to the 1995 data \cite{LEPEWWG9502}
occurred in $\Rb$ and $\Rc$. $\Rb$ is now given by 
$\Rb^{\exp} = 0.2178\pm 0.0011$ 
being hardly two standard deviations above the SM prediction for $\Rb$
for $\Mt=175\GeV$, 
while $\Rc^{\exp}$ is now in perfect agreement with the SM. We recall
that owing to its relatively large error $\Rc^{\exp}$ only
marginally influences the fits within the SM.
We expect that the above-mentioned
downward shift of $\Mt$, occurring as a consequence of including
$\Rb^{\exp}$ in the set of data being fitted, will be less drastic
in the fit based on the 1996 value of $\Rb^{\exp}$.
In addition to the change in $\Rb^{\exp}$, the Tevatron result for
the top-quark mass has changed from the 1995 value of 
$\Mt^{\exp}=180\pm 12\GeV$ \cite{mt95} to the 1996 value of 
$\Mt^{\exp}=175\pm 6\GeV$ \cite{mt96}. 
As a consequence of both, the lower value of $\Rb^{\exp}$ and the more
precise value of $\Mt^{\exp}$, we expect a very much reduced sensitivity
of the results for $\MH$ on whether $\Rb^{\exp}$ is included in or excluded 
from the fit.
Since the experimental values of other
observables have not changed very much, we expect that the
result for $\MH$ to be obtained from the full set of 1996 data will be
close to the values \refeq{eq:MH95}.

In what follows, we present a detailed analysis, in order to
investigate in how far the above expectation is actually reflected
in the results obtained from the 1996 data.

The results of a four-parameter $(\Mt,\MH,\alpz,\alpsz)$ fit
to the 1996 data, which are taken from 
\citere{LEPEWWG9602} and explicitly listed in \refta{tab:data},
are shown in the $(\MH,\Delta\chi^2)$ plots of \reffi{fig:Dchi}
and in \refta{tab:fit}.%
\footnote{
Fits of the Higgs-boson mass to the 1996 set of data were also 
performed in \citeres{LEPEWWG9602,ho96}. 
The results corresponding to the ones of \refta{tab:fit}, as far as
available, are in good agreement with ours.}
In this fit, $\MH$ and $\alpsz$ are treated as free fit parameters,
while for $\Mt$ and $\alpz$ the experimental information (central value
and error) is taken into account.
The reduced error in $\Mt^{\exp}=175\pm 6\GeV$, 
in comparison with the results in \citere{zph4},
leads to a steeper $\Delta\chi^2$ distribution in \reffi{fig:Dchi} and a
corresponding reduction of the errors on $\MH$ in \refta{tab:fit}.
We observe
that the results for $\MH$, as expected,
are indeed relatively independent of whether the value for $\Rb^{\exp}$
is or is not taken into account in the fits. In \reffi{fig:Dchi} and 
\refta{tab:fit},
we also present the results obtained from fits to the ``leptonic sector''
(see \refta{tab:data}), i.e.\ to the restricted set of data
consisting of $\swbar^2$, $\MW$, $\Gl$
accompanied by $\Mt^{\exp}$ and $\alpz$. 
It is noteworthy
that the results on $\MH$ from this restricted set of data
are as accurate as the results from the full
data sample, without being plagued, however, by potential
uncertainties related to $\Rb^{\exp}$ and $\alpsz$.
In fact, the quality of the fits in terms of $\chidof$ according
to \refta{tab:fit} in the case of the leptonic sector is 
better than in the case of the full set of data.

The central values of $\MH$ deduced from the full set of 1996 data
approximately resemble the values \refeq{eq:MH95} 
from the 1995 set with $\Rb^{\exp}$ excluded.
Indeed, in the case of $\swbar^2(\LEP+\SLD)$, the value of
$\MH=158^{+148}_{-84}\GeV$ resulting from the full 1996 set of data is
close to the value \refeq{eq:MH95} of $\MH=148^{+263}_{-103}\GeV$
\cite{zph4} obtained upon excluding $\Rb^{\exp}$ from the 1995 set of
data, while both values in turn differ appreciably from the value of
$\MH=81^{+144}_{-52}\GeV$ \cite{zph4} as obtained from the full set of
1995 data.

Turning to the dependence of the central value and the upper bound on
$\MH$ on the input for $\swbar^2$, we note that according
to \refta{tab:data} the values of $\swbar^2(\LEP+\SLD)$ and
$\swbar^2(\LEP)$ are only one standard deviation apart from each other.
Replacing $\swbar^2(\LEP+\SLD)$ by the higher value of
$\swbar^2(\LEP)$ nevertheless 
leads to an appreciable increase in the central value for $\MH$
by $\sim 100\GeV$, while the upper $1\si$ limits on $\MH$
increase by $\sim 200\GeV$. 
We finally note that $\swbar^2(\SLD)$, which is approximately $3\si$
below $\swbar^2(\LEP)$, when taken by itself 
in conjunction with all other data implies
$\MH=14^{+25}_{-7}\GeV$, which seriously violates the lower bound of
$65\GeV$ \cite{gr95} from the direct Higgs-boson search at LEP.
Obviously the rather unclear empirical information
on $\swbar^2$,
despite the high precision of the measurements,
represents one of the main sources of uncertainty in $\MH$
fits at present. 

The delicate interplay of the experimental results for $\swbar^2$, 
$\Rb$ and $\Mt$ in constraining $\MH$ is visualized in the
two-parameter $(\Mt,\MH)$ fits shown in \reffi{fig:mtmhfit}.
The (moderately) enhanced value of $\Rb^{\exp}$ still has the tendency of
pulling down the fit values for $\Mt$ to lower
values than the ones obtained by the
Tevatron measurement. Addition
of the fairly precise value of $\Mt^{\exp}=175\pm 6\GeV$ 
practically removes the dependence on $\Rb^{\exp}$ 
in accordance with the results for $\MH$ already displayed in
\reffi{fig:Dchi} and \refta{tab:fit}.

The dependence of $\Mt$ and $\MH$
on the central value as well as the error of the input parameter $\alpz$
has been taken into account
by including it as a fit parameter in the four-parameter fits,
which lead to the results in \reffi{fig:Dchi} and \refta{tab:fit}.
As a deviation in $\alpz$ 
by, e.g., the amount of one standard deviation is by no means excluded,
it is instructive to study the effect of varying the (fixed) 
input value of $\alpz$ by one standard deviation,
as displayed in the second row of \reffi{fig:mtmhfit}.
We note that the fit results and the corresponding variations with
different $\alpz$ 
in the case of the leptonic sector, which 
are not shown in \reffi{fig:mtmhfit}, are very close to the ones
obtained for the full set of data without $\Rb^{\exp}$, as shown in 
\reffi{fig:mtmhfit} (first column, second row).
The fit results for $\alpsz$, which has been treated as free fit parameter in
the four-parameter fits, are 
consistent with $\alpsz=0.123\pm0.006$
resulting from an event-shape analysis at LEP \cite{be95}.
Nevertheless, it is instructive to inspect also the dependence of the
fit results for $\Mt$ and $\MH$ on a fixed input for $\alpsz$.
The effect of varying $\alpsz$ is illustrated in the third row
of \reffi{fig:mtmhfit}.

The last row of \reffi{fig:mtmhfit}
shows the dependence of the $1\si$ contours
in the $(\Mt,\MH)$ plane on the experimental
input for the effective weak mixing angle, $\swbar^2$. The level
of sensitivity reached by the data for $\swbar^2$ becomes apparent
by noting that the shift in $\MH$ resulting from replacing
$\swbar^2(\LEP+\SLD)$ by $\swbar^2(\LEP)$ is of roughly the same
magnitude as the shift due to the change in $\alpz$ by one
standard deviation shown in the second row of \reffi{fig:mtmhfit}.

The above discussion based on \refta{tab:fit} and 
\reffis{fig:Dchi} and \ref{fig:mtmhfit} may be
summarized by concluding that the $\Rb^{\exp}$ effect of the 1995
data of effectively pulling down the fit value for $\Mt$, and
consequently of $\MH$, is still present in the 1996 data, 
even though the magnitude of the effect is strongly reduced.
Nevertheless,
one notices that the
central values of $\MH$ obtained from the leptonic sector and 
based on the
full set of data upon excluding $\Rb^{\exp}$ are very close together,
while differing by 
$\sim 30\GeV$ from the result from the full
set of data. Taking into account the fact that the $1\si$ errors of
$\MH$ obtained in all three cases are approximately the same, while
$\chidof$ is best in the fit to the leptonic set of data, we prefer to
quote this result of
\beq
\begin{array}[b]{rcl}
\MH &=& 190^{+174}_{-102}\GeV, \qquad \mbox{using} \qquad
\parbox{7cm}{\hfill$\swbar^2(\LEP+\SLD)=0.23165\pm 0.00024,$} 
\\
\MH &=& 296^{+243}_{-143}\GeV, \qquad \mbox{using} \qquad
\parbox{7cm}{\hfill$\swbar^2(\LEP)=0.23200\pm 0.00027,$}
\end{array}
\label{eq:MH96}
\eeq
as our final one from the 1996 electroweak data. 
We stress again that \refeq{eq:MH96} has the advantage of being
independent of 
$\Rb^{\exp}$ and $\alpsz$. 
In comparison with the result \refeq{eq:MH95} based on the 1995 set of 
data, the upper 
$1\si$ bounds \refeq{eq:MH96} on $\MH$ based on the 1996 set of data
are strongly reduced to $\MH\lsim 360\GeV$ and
$\MH\lsim 540\GeV$, for $\swbar^2(\LEP+\SLD)$ and $\swbar^2(\LEP)$,
respectively.

{\doublerulesep 3pt
\btab
\bce
\begin{tabular}{|@{}c@{}||c|c|}
\hline
leptonic sector & \multicolumn{2}{c|}{hadronic sector} \\
\hline
\hline
$\Gl = 83.91 \pm 0.11 \MeV$ & 
$R = 20.778 \pm 0.029$ & $\GT = 2494.6 \pm 2.7 \MeV$ \\
\hline
$\swbar^2 (\LEP) = 0.23200 \pm 0.00027$ & 
$\si_{\mathrm h} = 41.508 \pm 0.056$ &
$\Gh = 1743.6 \pm 2.5 \MeV$ \\
\hline
$\swbar^2 (\SLD) = 0.23061 \pm 0.00047$ & 
$\Rb = 0.2179 \pm 0.0012$ &
$\Gb = 379.9 \pm 2.2 \MeV$ \\
\hline
$\swbar^2 (\LEP+\SLD) = 0.23165\pm0.00024$ &
$\Rc = 0.1715 \pm 0.0056$ &
$\Gc = 299.0 \pm 9.8 \MeV$ \\ \hline
$\MW = 80.356 \pm 0.125 \GeV$ & \multicolumn{2}{c|}{} \\
\hline \hline
input parameters & \multicolumn{2}{c|}{correlation matrices} \\
\hline \hline
\begin{array}[b]{c}
\MZ = 91.1863 \pm 0.0020 \GeV \\ \hline
\hspace{20pt} \GF = 1.16639 (2) \cdot 10^{-5} \GeV^{-2} \hspace{20pt} \\ \hline
\alpz^{-1} = 128.89 \pm 0.09 \\ \hline
\alpsz = 0.123\pm0.006 \\ \hline
\Mb = 4.7\GeV \hspace{1pt} \\ \hline
\Mt = 175 \pm 6 \GeV\\
\end{array}
& \multicolumn{2}{c|}{
\begin{array}[b]{|c||c|c|c|c|}
\hline
& \si_{\mathrm h} & R & \GT \\ \hline\hline
\si_{\mathrm h} & \phantom{-}1.00 & \phantom{-}0.15 & -0.14 \\ \hline
R               & \phantom{-}0.15 & \phantom{-}1.00 & -0.01 \\ \hline
\GT             & -0.14 & -0.01 & \phantom{-}1.00 \\ \hline
\earr
\qquad
\begin{array}[b]{|c||c|c|c|}
\hline
& \Rb & \Rc \\ \hline\hline
\Rb  & \phantom{-}1.00 & -0.23 \\ \hline
\Rc  & -0.23 & \phantom{-}1.00 \\ \hline
\earr} \\ \hline
\end{tabular}
\ece
\caption[]{
The precision data used in the fits, consisting of the LEP data
\cite{LEPEWWG9602}, the SLD value \cite{SLD} for $\swbar^2$, and the world 
average \cite{MW} for $\MW$.
The partial widths $\Gl$, $\Gh$, $\Gb$, and $\Gc$ are obtained from the
observables $R = \Gh/\Gl$, $\si_{\mathrm h} =
(12\pi\Gl\Gh)/(\MZ^2\Ga^2_{\mathrm T})$, 
$\Rb =  \Gb/\Gh$,
$\Rc =  \Gc/\Gh$, and  $\GT$ using the given correlation
matrices. The data in the upper left-hand column will be referred to as
``leptonic sector'' in the fits. Inclusion of the data in the upper
right-hand column will be referred to as fitting ``all data''.
If not stated otherwise,
the theoretical predictions are based on the input parameters
given in the lower left-hand column of the table, where $\alpz$ is taken
from \citere{bu95}, $\alpsz$ results from the event-shape 
analysis~\cite{be95} at LEP, and $\Mt$ represents the direct
Tevatron measurement~\cite{mt96}.}
\label{tab:data}
\etab
}

\begin{figure}
\begin{center}
\begin{picture}(15,13.0)
\put(-2.5,-13.3){\includegraphics{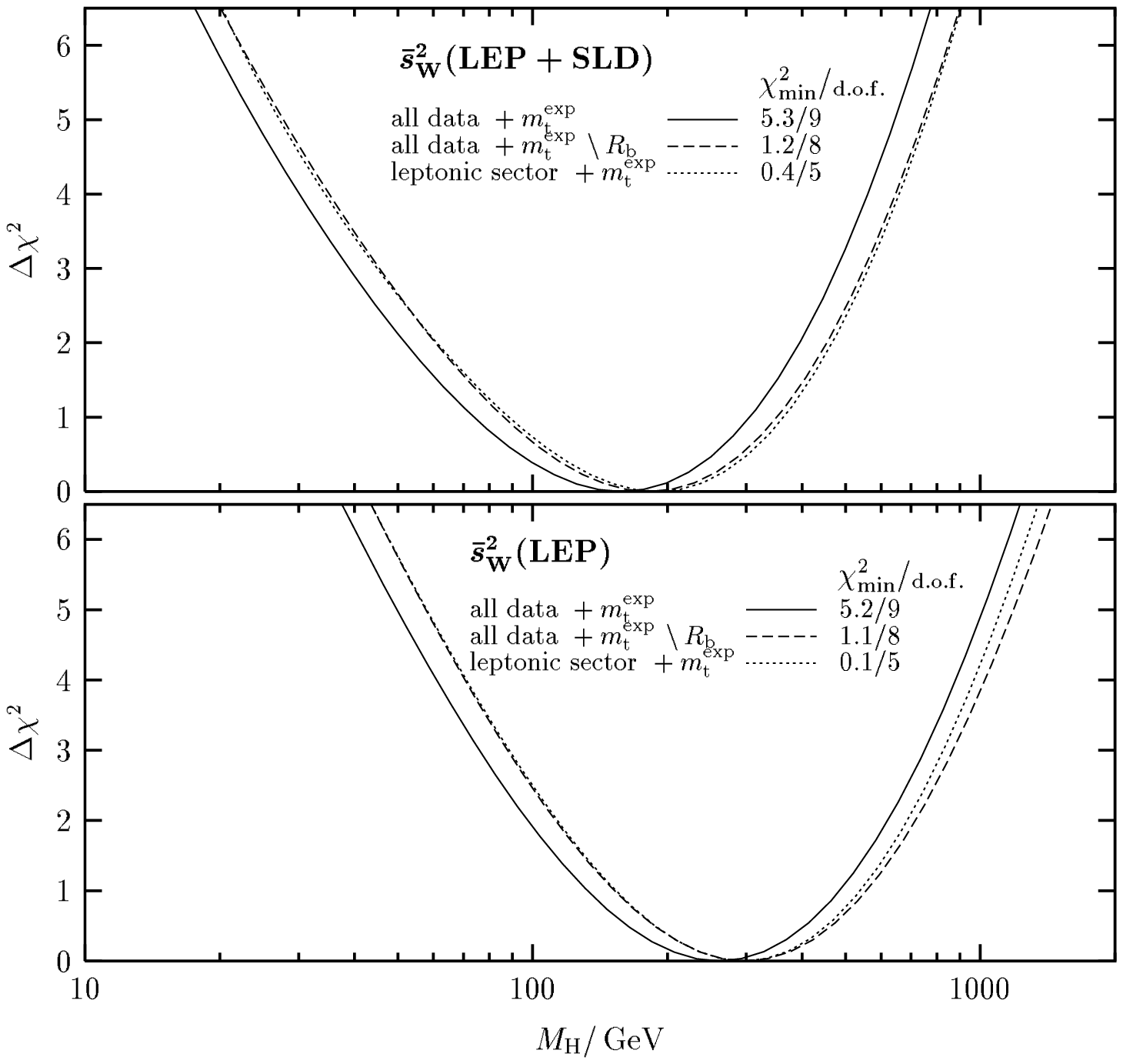}}
\end{picture}
\end{center}
\caption{$\De\chi^2=\chi^2-\chi^2_{\mathrm{min}}$ is plotted against
$\MH$ for the $(\Mt,\MH,\alpz,\alpsz)$ fit to various sets of 
physical observables.
For a chosen input for
$\swbar^2$, as indicated, we show the result of a fit to
\protect\\
(i) the full set of 1996 data, 
$\swbar^2$, $\MW$, $\GT$, $\si_{\mathrm{h}}$, $R$, $\Rb$, $\Rc$, 
together with
$\Mt^{\exp}$, $\alpz$,
\protect\\
(ii) the 1996 set of (i) upon exclusion of $\Rb$,
\protect\\
(iii) the 1996 ``leptonic sector'' of $\swbar^2$, $\MW$, $\Gl$, 
together with $\Mt^{\exp}$, $\alpz$.
}
\label{fig:Dchi}
\efi

\btab
\bce
\begin{tabular}{|l|l||c|c|c|c|}
\hline
& data & $\Mt$ & $\MH$ & $\alpsz$ & $\chidof$ \\
\hline\hline
$\swbar^2(\LEP+\SLD)$ & '96 & 
$172^{+6}_{-6}$ & $158^{+148}_{-84}$ & $0.121^{+0.003}_{-0.003}$ & $5.3/9$ \\
\cline{2-6}
& '96 $\backslash\Rb$ & 
$174^{+6}_{-6}$ & $182^{+169}_{-96}$ & $0.121^{+0.003}_{-0.003}$ & $1.2/8$ \\
\cline{2-6}
& '96 {leptonic sector} & 
$174^{+6}_{-6}$ & $190^{+174}_{-102}$ & & $0.4/5$ \\
\cline{2-6}\cline{2-6}
& '95 $\backslash\Rb$ & 
$175^{+12}_{-11}$ & $148^{+263}_{-103}$ & $0.122^{+0.004}_{-0.004}$ & $7.3/8$ \\
\hline\hline
$\swbar^2(\LEP)$ & '96 & 
$173^{+6}_{-6}$ & $261^{+224}_{-128}$ & $0.123^{+0.003}_{-0.003}$ & $5.2/9$ \\
\cline{2-6}
& '96 $\backslash\Rb$ & 
$175^{+6}_{-6}$ & $299^{+258}_{-147}$ & $0.123^{+0.003}_{-0.003}$ & $1.1/8$ \\
\cline{2-6}
& '96 {leptonic sector} & 
$175^{+6}_{-6}$ & $296^{+243}_{-143}$ & & $0.1/5$ \\
\cline{2-6} 
& '95 $\backslash\Rb$ & 
$178^{+12}_{-12}$ & $343^{+523}_{-219}$ & $0.124^{+0.004}_{-0.004}$ & $6.6/8$ \\
\hline
\end{tabular}
\ece
\caption{
The results of the $(\Mt,\MH,\alpz,\alpsz)$
fits to various sets of experimental data, as indicated. In detail,
the rows (i) to (iv) are based on:
\protect\\
(i) the full 1996 set of 
$\swbar^2$, $\MW$, $\GT$, $\si_{\mathrm{h}}$, $R$, $\Rb$, $\Rc$, 
together with
$\Mt^{\exp}$, $\alpz$,
\protect\\
(ii) the set as in (i) but excluding $\Rb$,
\protect\\
(iii) the 1996 ``leptonic sector'' of $\swbar^2$, $\MW$, $\Gl$, 
together with
$\Mt^{\exp}$, $\alpz$,
\protect\\
(iv) the 1995 set as in (i), but excluding $\Rb$,
\protect\\
where for $\swbar^2$ the values of $\swbar^2(\LEP+\SLD)$ and
$\swbar^2(\LEP)$ are used in the fits, as indicated.
The values for $\alpz$ obtained in the fits reproduce the input
given in \refta{tab:data} and accordingly are suppressed in the present 
table.
For the leptonic sector $\alpsz=0.123$ is used as fixed input (needed
only for two-loop corrections), whereas
$\alpsz$ is treated as unconstrained fit parameter in the other fits.}
\label{tab:fit}
\etab

\begin{figure}
\caption[xxx]{
The results of the two-parameter $(\Mt,\MH)$ fits within the SM
are displayed in the $(\Mt, \MH)$ plane. The different columns refer 
to the sets of experimental data used in the corresponding 
fits, \protect\\
(i) ``all data $\backslash\Rb$'': $\swbar^2(\LEP+\SLD)$, 
$\MW$, $\GT$, $\si_{\mathrm{h}}$, $R$, $\Rc$, 
\protect\\
(ii) ``all data'': $\Rb$ is added to set (i), 
\protect\\
(iii) ``all data + $\Mt^{\exp}$'': $\Rb,\Mt^{\exp}$ are added to the set (i).
\protect\\
The second and third row shows the shift resulting from changing 
$\alpz^{-1}$ and $\alpsz$, respectively,
by one standard deviation in the SM prediction.
The fourth row shows the effect of replacing $\swbar^2(\LEP+\SLD)$ by 
$\swbar^2(\LEP)$ and $\swbar^2(\SLD)$ in the fits.
Note that the $1\sigma$ boundaries given in the first row 
are repeated identically in each row, in order to facilitate comparison with 
other boundaries. 
The value of $\chidof$ given in the plots refers to the 
central values of $\alpz^{-1}$ and $\alpsz$.
In all plots the empirical value 
of $\Mt^{\exp} = 175 \pm 6 \GeV$ is also indicated.}
\label{fig:mtmhfit}
\efi

\addtocounter{figure}{-1}
\begin{figure}
\begin{center}
\begin{picture}(16,21)
\put(-0.2,-0.7){\includegraphics{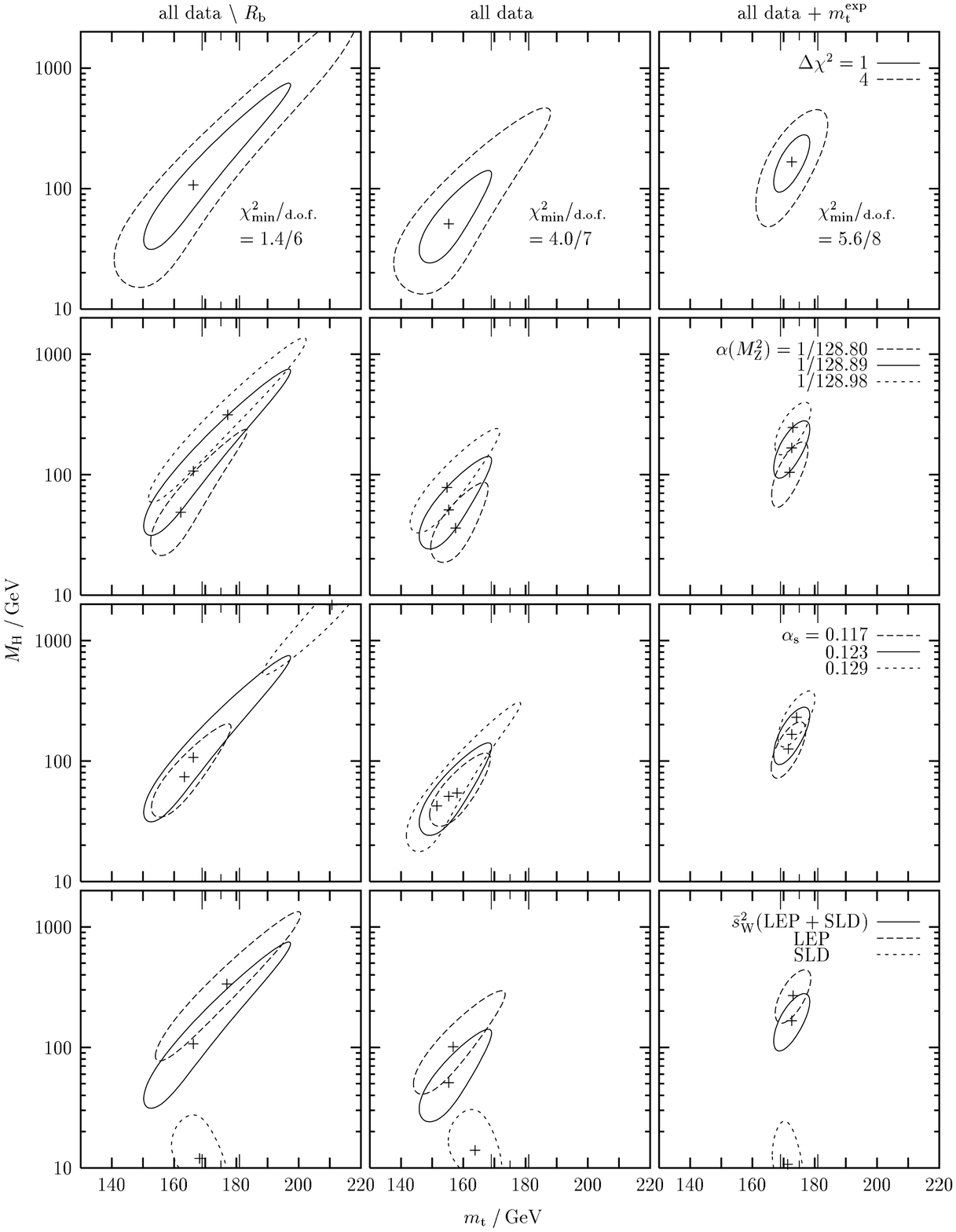}}
\end{picture}
\end{center}
\caption{}
\efi

\clearpage

\section*{Analysis in terms of effective parameters}

We turn to the second part of the present work and investigate
electroweak interactions from the point of view of 
the effective
Lagrangian developed in \citeres{zph1,zph2,kn91,bi93}. 
This approach allows for a detailed
assessment of which elements of the electroweak theory are quantitatively
tested by the precision experiments at the Z-boson resonance. 
In particular, the question in how far the non-Abelian structure
of the theory enters the predictions 
(and in how far it is tested)
and questions on the empirical evidence for the Higgs mechanism
may be answered by this approach.

In addition to the canonical input parameters---the Fermi coupling $\GF$,
the effective electromagnetic coupling at the Z-boson mass scale,
$\alpz$, and the Z-boson mass, 
$\MZ$---for the interactions of the
vector bosons with charged leptons the effective Lagrangian contains
three parameters, $\De x$, $\De y$, $\eps$.
The parameters $\De x$, $\De y$, $\eps$
are related to SU(2) breaking in the 
vector-boson masses, in their
couplings to charged leptons, and in the mixing among the neutral
vector bosons, respectively. 
For vanishing values of $\De x$, $\De y$, $\eps$, 
in the leptonic sector the Lagrangian coincides with the one of the SM in the
unitary gauge, and its (tree-level) predictions
coincide with the $\alpz$-Born approximation of the SM.
In order to extend the Lagrangian to the
interactions of neutrinos and quarks, additional parameters,
related to the corresponding couplings, have been introduced
in \citere{zph2}.
They are
given by $\De y_\nu$ for the neutrino, $\De\yb$ for the bottom quark,
and $\De\yh$ for the remaining light quarks.

In our analysis we restrict ourselves to a four-parameter
($\De x,\De y,\eps,\De\yb)$ fit, 
assuming that $\De y_\nu$ and $\De\yh$ are
well represented by vertex corrections in the SM.
We note in passing 
that the corrections entering $\De y_\nu$ and $\De\yh$ only
depend on vector-boson--fermion interactions and do not involve the
non-Abelian structure of the theory (see \citere{zph2}).
The results of our fit to the data of \refta{tab:data} 
are shown in \refta{tab:xyeb} and \reffi{fig:xyeb}.

The comparison in \refta{tab:xyeb} with the results of the analysis 
in \citere{zph2}, which have been  based on
the 1994 data, shows a significant decrease of the experimental errors
to roughly two thirds of their 1994 value. The most drastic shift in
the central values of the parameters occurred in $\De\yb$, due to
the shift of the experimental value for $\Rb$, as discussed
in the preceding section. The $1\sigma$ ellipse in the 
$(\De\yb,\eps)$ plane of \reffi{fig:xyeb} now includes the value of 
$\Mt= 160\GeV$,
while with the 1994 value it only reached values of 
$\Mt\sim 100\GeV$%
\footnote{When comparing the present results with the ones of
\citere{zph2}, one should notice that in \citere{zph2} the value of
$\alpsz=0.118\pm0.007$ was used and that the ellipses shown in the
plots for the effective parameters were scaled by a factor $1.4$
(corresponding to a confidence level of 61\% but not 83\%, as
erroneously indicated in \citere{zph2}).}.

Otherwise, the present analysis at an improved level of accuracy
confirms conclusions drawn previously \cite{zph1,zph2}:
\begin{enumerate}
\item
The effective parameters deviate significantly from 
zero, the value which corresponds to the $\alpz$-Born approximation 
of the SM. Genuinely electroweak corrections are definitely
required within the SU(2)$\times$U(1) theory to obtain consistency
with the experimental results%
\footnote
{Prior to the start of LEP experiments it was stressed~\cite{go88} that
the difference between 
the dominant fermion-loop and the full one-loop corrections
sets the scale for genuine electroweak precision tests.
This precision was 
first reached in 1994 \cite{zph1} but not yet in 1993 \cite{bi93,no93}.}.
\item
In the case of the ``mass parameter'' $\De x$ and the ``mixing parameter''
$\eps$, the pure contributions of the fermion loops to the \PWpm-, 
Z-, and $\gamma$-boson propagators, i.e.\ neglecting all contributions
involving loops containing vector bosons,
\beq
\De x\simeq\De x_{\fer}, \qquad \eps\simeq\eps_{\fer},
\label{eq:Fxe}
\eeq
lead to consistency with the experimental data. Note that $\De x_\fer$
contains a dominating $\Mt^2$ and a $\log\Mt$ term besides a small
constant contribution, while 
$\eps_\fer$ is dominated by a large constant term
and contains an additional $\log\Mt$ contribution. The 
(logarithmic)
Higgs-boson mass dependence is entirely contained in the bosonic 
contributions,
$\De x_{\bos}$ and $\eps_{\bos}$, to $\De x$ and $\eps$. In view of the
results of the $\MH$ fits discussed above, it is not
surprising that these bosonic contributions are not well resolved 
by the projection of the ellipsoid into the $(\De x,\eps)$ plane
shown in \reffi{fig:xyeb}. 
The shift to somewhat higher values
of $\MH$, if $\swbar^2(\LEP+\SLD)$ is replaced 
by $\swbar^2(\LEP)$,
or if $\alpz^{-1}$ is replaced by $\alpz^{-1}+\de\alpz^{-1}$,
is nevertheless clearly visible in \reffi{fig:xyeb}.
Note that both 
the uncertainty in the experimental value of $\swbar^2$ as well as the
one in $\alpz$
mainly affect the mixing parameter $\eps$.
\item
In the ``coupling parameter'' $\De y$ a contribution beyond fermion
loops,
\beq
\De y = \De y_{\fer} + \De y_{\bos},
\label{dy}
\eeq
is definitely required for
consistency with the data. As seen in \refta{tab:xyeb}, the large negative
fermion-loop contribution in the SM
is overcompensated by a positive bosonic
part which is practically independent of $\MH$. As pointed out in 
\citere{zph3}, the large fermionic and bosonic
contributions to $\De y$ may be traced back to the use of the
low-energy parameter $\GF$ as input for the theoretical predictions.
Indeed, the main fermionic and bosonic contributions to $\De y$ are
identical to the loop contributions $\De y^{\SC}$, connecting the
low-energy charged-current coupling, $g_{\PWpm}^2(0)\propto\GF\MW^2$,
with the high-energy charged-current coupling,
$g_{\PWpm}^2(\MW^2)\propto\GF\MW^2/(1+\De y^{\SC})$, appearing in
the leptonic width of the W~boson.
For details, we refer to \citere{zph3}. 
Even though the theoretical 
value of $\De y$ is extremely insensitive to
$\MH$ and may even be derived in the framework of a massive 
vector-boson theory \cite{di95}, the agreement 
of the theoretical prediction with experiment constitutes 
significant positive empirical evidence for
the non-Abelian sector of the SM.
\end{enumerate}

We finally note that the isolation of the experimentally resolved dominant,
$\MH$-in\-sen\-si\-tive
bosonic corrections (in $\De y$)
from the small, $\MH$-dependent ones (in $\De x$ and $\eps$) is a specific
feature of our choice of parameters, naturally implied by examining
SU(2) breaking in an effective Lagrangian. 
This separation is not present in
the widely used $\eps_i$ parameters~\cite{al93} related to ours 
via~\cite{zph2}
\beqar
\parbox{6cm}{$\eps_1=\De x-\De y+0.2\times 10^{-3},$} &&
\eps_2=-\De y+0.1\times 10^{-3}, \nn\\
\parbox{6cm}{$\eps_3=-\eps+0.2\times 10^{-3},$} && 
\eps_\Pb=-\De\yb/2-0.1\times 10^{-3}.
\eeqar
In $\eps_1$, the $\MH$-dependent 
contribution to the mass parameter,
$\De x$, appears in linear combination with the $\MH$-insensitive
bosonic correction contained in the coupling parameter $\De y$.

\btab
\begin{tabular}{|r||r|r||r|}
\hline
\raisebox{-1em}[0cm][0cm]{\refta{tab:xyeb}a}
& \multicolumn{2}{c||}{'96 data} 
& \multicolumn{1}{c|}{'94 data} \\
\cline{2-4}
& \multicolumn{1}{c|}{$\swbar^2(\LEP+\SLD)$} 
& \multicolumn{1}{c||}{$\swbar^2(\LEP)$} 
& \multicolumn{1}{c|}{$\swbar^2(\LEP)$} \\
\hline\hline
$\De x/10^{-3}$ & 
$11.8\pm3.2\pm0.2\mp0.1$ &  
$12.6\pm3.2\pm0.2\mp0.0$ & 
 $9.6\pm4.7\pm0.2\mp0.0$ \\
\hline
$\De y/10^{-3}$ &  
$8.4\pm3.3\pm0.2\pm0.4$ &  
$8.9\pm3.3\pm0.2\pm0.4$ &  
$5.6\pm4.8\pm0.2\pm0.4$ \\
\hline
$\eps/10^{-3}$ & 
$-3.8\pm1.2\mp0.5\pm0.3$ & 
$-4.9\pm1.2\mp0.5\pm0.3$ & 
$-5.2\pm1.8\mp0.7\pm0.3$ \\
\hline
$\De\yb/10^{-3}$ &  
 $6.7\pm3.8\pm0.0\pm4.4$ & 
 $6.1\pm3.8\pm0.0\pm4.4$ &  
$-3.3\pm5.9\pm0.0\pm5.8$ \\
\hline
\end{tabular}
\\[.8em]
\begin{tabular}{|r||r|r|}
\hline
\raisebox{-1em}[0cm][0cm]{\refta{tab:xyeb}b}
& \multicolumn{2}{c|}{theory} \\
\cline{2-3}
& \multicolumn{1}{c|}{full} & ferm.~1-loop \\
\hline\hline
 $\De x/10^{-3}$ & $11.8^{+0.7-1.4}_{-0.7+1.1}$ & $11.7^{+0.8}_{-0.8}\quad$ 
\\ \hline
 $\De y/10^{-3}$ & $7.3^{+0.1-0.0}_{-0.1+0.2}$ & $-6.4^{+0.1}_{-0.1}\quad$ 
\\ \hline
  $\eps/10^{-3}$ & $-5.2^{+0.0-0.6}_{-0.0+1.3}$ & $-5.8^{+0.0}_{-0.0}\quad$ 
\\ \hline
$\De\yb/10^{-3}$ & $11.8^{+1.1+0.1}_{-1.0+0.1}$& $0\phantom{^{+0.0}}\quad$ 
\\ \hline
\end{tabular}
\caption{
The results for the four parameters $\De x$, $\De y$, $\eps$, $\De\yb$,
\protect\\
(a) obtained by a fit to the experimental data, as specified (the 1994
results are taken from \citere{zph2}).
The first error is statistical ($1\si$), the second due to
the deviation by replacing $\alpz^{-1}\to\alpz^{-1}\pm\de\alpz^{-1}$,
and the third due to $\alpsz\to\alpsz\pm\de\alpsz$.
\protect\\
(b) as predicted within the SM for $\Mt= 175\GeV$ and $\MH=300\GeV$.
The ``full'' predictions, which include the complete one-loop 
and leading two-loop corrections (see \citere{zph2}),
are compared to the 
${\cal O}(\alpha)$ 
fermion-loop contributions. The first
upper/lower deviations correspond to $\pm 6\GeV$ in $\Mt$, the second
(if given) to the variations of $\MH$ from $60\GeV$ (lower) to $1\TeV$
(upper).}
\label{tab:xyeb}
\etab

\begin{figure}
\begin{center}
\begin{picture}(16,15.5)
\put(-2.0,-4.7){\includegraphics{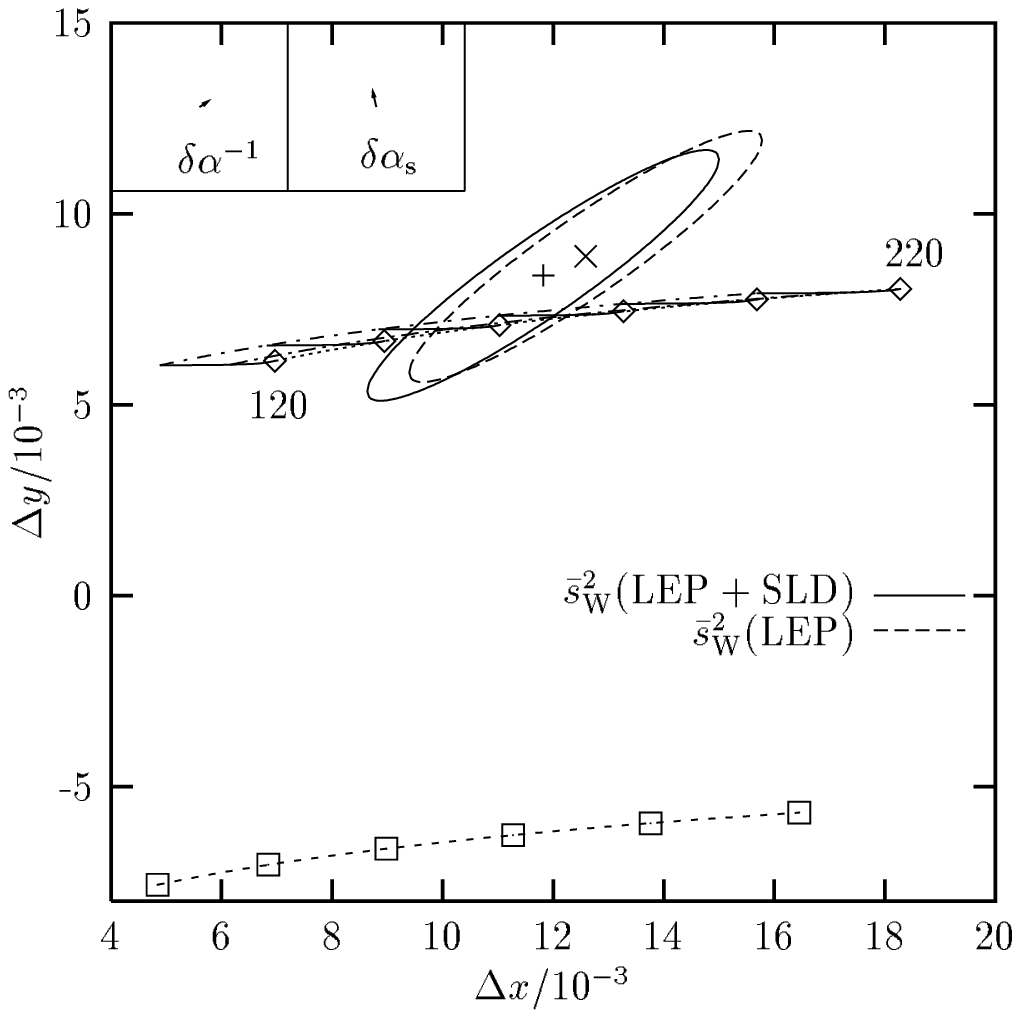}}
\put( 6.0,-4.7){\includegraphics{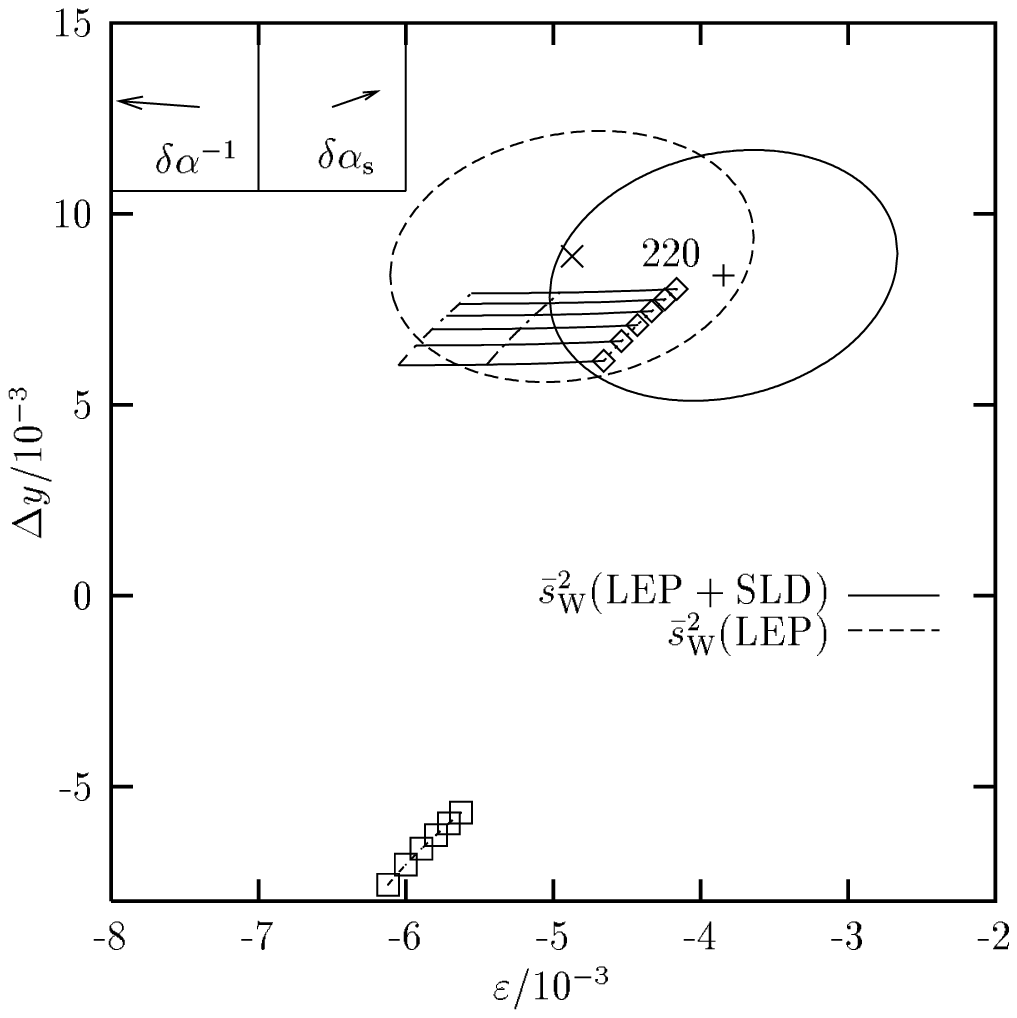}}
\put(-2.0,-12.7){\includegraphics{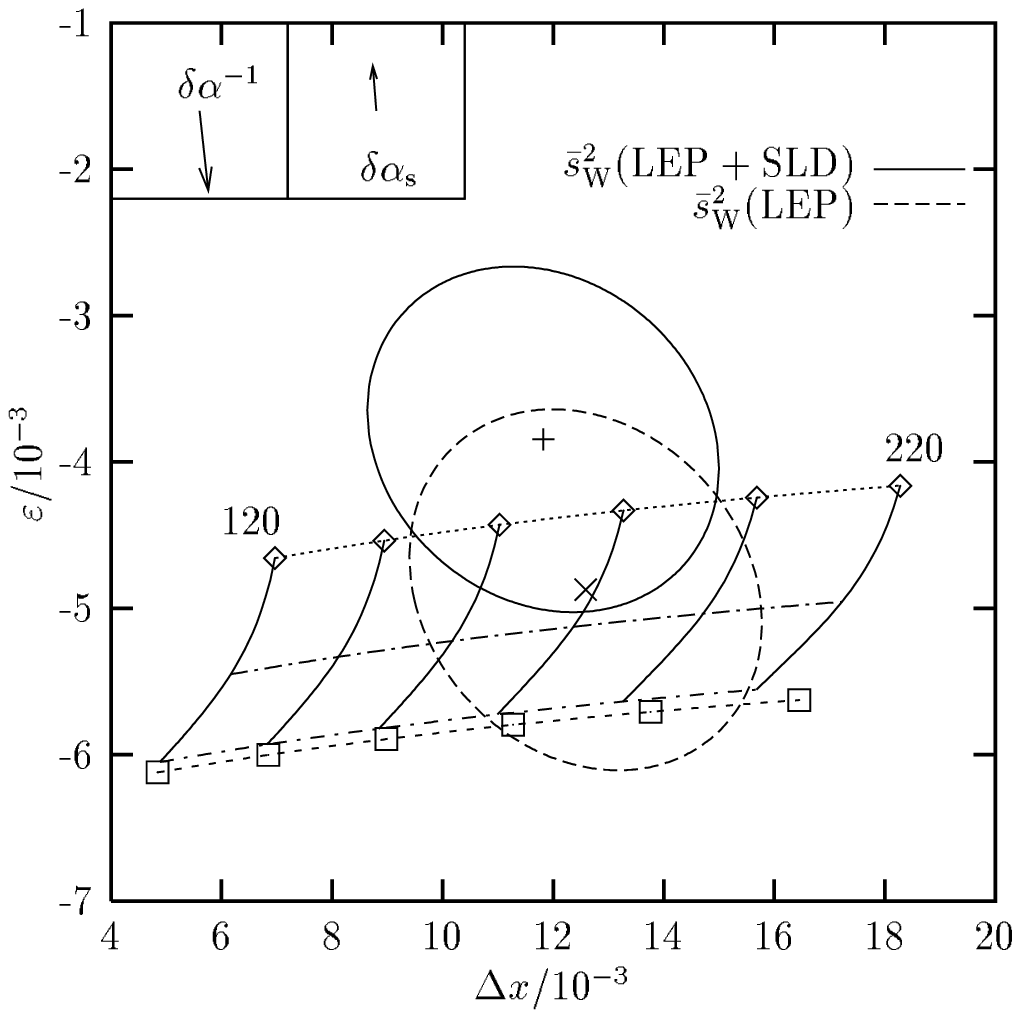}}
\put( 6.0,-12.7){\includegraphics{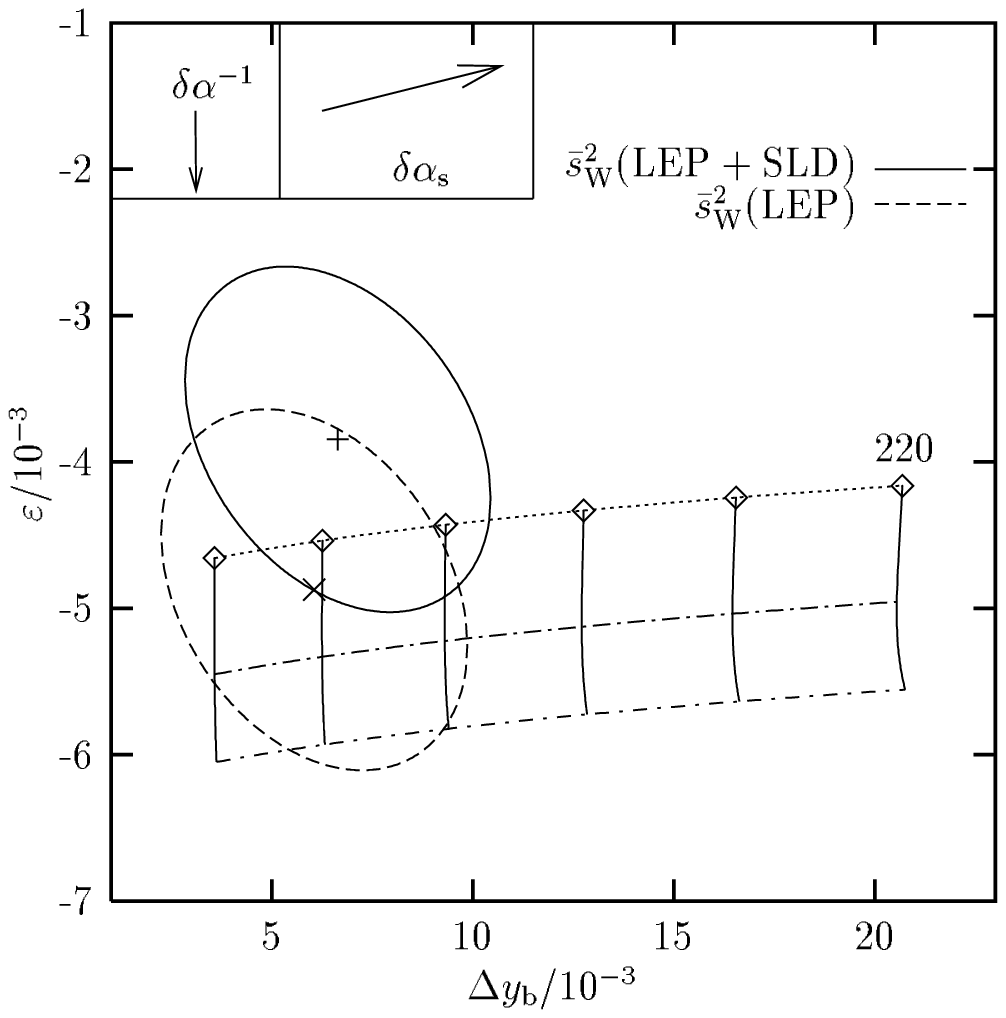}}
\end{picture}
\end{center}
\caption{
The projections of the $1\si$ ellipsoid of 
the electroweak parameters $\De x$, $\De y$, $\eps$,
$\De\yb$ obtained from the 1996 set of data in comparison
with the SM predictions. Both the
results obtained from using $\swbar^2(\LEP)$ and $\swbar^2(\LEP+\SLD)$
as experimental input are shown. The full SM predictions
correspond to Higgs-boson masses of $100\GeV$ (dotted with diamonds), 
$300\GeV$ (long-dashed dotted) and $1\TeV$ (short-dashed dotted)
parametrized by the top-quark mass ranging from
$120\GeV$ to $220\GeV$ in steps of $20\GeV$. 
The pure fermion-loop
prediction is also shown (short-dashed curve with squares) for the
same values of $\Mt$.
The arrows indicate the shifts of the centres of the ellipses upon
changing $\alpz^{-1}$ to $\alpz^{-1}+\de\alpz^{-1}$ and $\alpsz$ to
$\alpsz+\de\alpsz$.}
\label{fig:xyeb}
\efi

\clearpage

\section*{Conclusions}

With the 1996 data, the agreement between experiment and the predictions
of the electroweak Standard Model (SM) has become even more impressive.
From the analysis in terms of the parameters in an effective Lagrangian
we learned that the non-Abelian structure of the theory entering the
bosonic loops is quantitatively supported by the empirical data. The
upper $1\si$ bounds on the Higgs-boson mass, $\MH$, have improved to 
$360\GeV$ and $540\GeV$ based on $\swbar^2(\LEP+\SLD)$ and
$\swbar^2(\LEP)$, respectively. Changing $\alpz$ by one standard 
deviation leads to changes in the upper bounds on $\MH$
which are similar to the ones induced by the difference between
$\swbar^2(\LEP+\SLD)$ and $\swbar^2(\LEP)$.
At the 95\% C.L.\ we obtain $\MH\lsim 550\GeV$ and $\MH\lsim 800\GeV$,
respectively. We stress that these bounds already follow from the
reduced set of data of $\swbar^2$, $\MW$, $\Gl$, $\Mt^{\exp}$, and $\alpz$, 
thus avoiding potential uncertainties related to $\Rb$, $\Rc$, and $\alpsz$.
Moreover, the bounds on $\MH$ are not
significantly further improved by including also the experimental
information on the inclusive (hadronic and total) Z-boson decays.
The fact that the bounds on $\MH$ lie in the perturbative regime of the 
SM may be interpreted as supporting the concept of the Higgs mechanism,
even though the ultimate answer to the question of its realization in
nature cannot be given as long as experimental evidence for the
existence of the Higgs boson is missing.


\end{document}